\documentclass[11pt]{article}

\usepackage[margin=1in]{geometry}
\usepackage[T1]{fontenc}
\usepackage{lmodern}
\usepackage{microtype}
\usepackage{amsmath}
\usepackage{amssymb}
\usepackage{amsthm}
\usepackage{booktabs}

\newtheorem{definition}{Definition}

\newtheorem{property}{Property}

\usepackage{enumitem}
\usepackage{xcolor}
\usepackage[hidelinks,colorlinks=true,linkcolor=blue!50!black,
            urlcolor=blue!50!black,citecolor=blue!50!black]{hyperref}
\usepackage[capitalize,noabbrev]{cleveref}
\crefname{property}{Property}{Properties}
\Crefname{property}{Property}{Properties}
\usepackage[numbers,sort&compress]{natbib}
\usepackage{listings}
\usepackage{tikz}
\usetikzlibrary{arrows.meta,positioning,fit,backgrounds}

\newcommand{\wellknown}{\texttt{/.well-known/enclawed-clearance.json}}

\title{\bfseries Attested Tool-Server Admission:\\
       A Security Extension to the\\
       Model Context Protocol}

\author{%
  Alfredo Metere\\
  Enclawed LLC\\
  \texttt{alfredo.metere@enclawed.com}
}
\date{\today\\[4pt]{\small Preprint: \href{https://arxiv.org/abs/2605.24248}{arXiv:2605.24248} $\,\cdot\,$ Archived record: \href{https://doi.org/10.5281/zenodo.20349263}{doi:10.5281/zenodo.20349263}}}

\begin{document}

\maketitle

\begin{abstract}\noindent
The Model Context Protocol (MCP) standardizes how a large-language-model (LLM)
agent and an external tool server exchange messages, but not trust: a host reads
a server's self-declared tool list and dispatches calls, with no notion of which
servers it may use, at what sensitivity, or which of a server's tools are in
bounds. This work grew out of a concrete need --- letting the Enclawed agent use
Google's externally-operated MCP servers (Gmail, Calendar, Drive) safely,
admitting the server and bounding the tools it may drive, without changing MCP or
Enclawed's own tool application-programming interface (API). The mechanism we
built, \texttt{mcp-attested} (shipped in both the open \texttt{enclawed-oss}
distribution and the enclaved flavor), generalizes: the gap that makes an
unmediated third-party connection unsafe for one user makes a regulated
deployment impossible to accredit. We close it with three additive mechanisms:
(1) a small, offline-signed \emph{clearance assertion} a server publishes at a
well-known Uniform Resource Identifier (URI) and a host verifies against a pinned
trust root before any tool dispatch; (2) a \emph{deny-by-default per-server tool
allowlist}, so admitting a server is not trusting its every tool; and (3) a
\emph{flavor-gated} enforcement mode that turns the checks from warnings into
hard denials, with every decision written to a tamper-evident audit log. We give
the wire format, the verification algorithm, a security analysis, and an
LLM-driven adversarial evaluation; we then state the design in normative
Request-for-Comments (RFC 2119) form --- schema, verification rules, error
registry, well-known registration, and machine-checkable conformance vectors ---
so it can be adopted as an MCP addendum rather than reinvented. An unextended
host ignores the well-known document and behaves exactly as today.
\end{abstract}

\section{Introduction}

The Model Context Protocol (MCP)~\cite{mcp-spec,mcp-intro} standardized
something the agent ecosystem badly needed: a single way for an LLM-driven
host to enumerate the tools, resources, and prompts that an external server
offers, and to invoke them over a uniform JSON-RPC (JavaScript Object Notation remote procedure call)~\cite{jsonrpc} envelope.
Its adoption was rapid precisely because it is minimal. A host connects to a
server over a transport (a subprocess pipe, \texttt{stdio}; the HyperText
Transfer Protocol, HTTP; or a streaming variant), issues
\texttt{tools/list}, receives a list of tool descriptors, and thereafter
issues \texttt{tools/call} with a tool name and an arguments object.

\paragraph{Where this work began.}
We wanted the Enclawed agent to use Google's externally-operated Workspace MCP
servers (Gmail, Calendar, Drive) as ordinary tools. Wiring them in the obvious
way trusts whatever tool list the server advertises and hands the model the
server's \emph{entire} surface --- including, say, a ``delete every message''
tool --- the moment it connects. We wanted instead to \emph{admit} such a server
deliberately and \emph{bound} the tools the agent can drive, without changing MCP
or Enclawed's own tool API. The component built for that, \texttt{mcp-attested},
generalizes into the extension this paper proposes; it ships in both the open
\texttt{enclawed-oss} distribution and the enclaved flavor.

That minimalism is the security gap. MCP is silent on \emph{whether} a host
should talk to a given server, \emph{what sensitivity} may cross the connection,
and \emph{which subset} of the server's advertised tools the host may drive; the
tool list and the server's identity --- beyond whatever its Transport Layer
Security (TLS) certificate asserts --- are taken on faith. The result is a
confused deputy~\cite{confused-deputy}: an LLM steered by adversarial
context~\cite{owasp-llm-top-10,mitre-atlas} calls a tool, and nothing sits
between that decision and the network write.

An unmediated connection is unsafe in \emph{every} deployment: the operator
chose the server, but not the tool surface the model can reach, and a
prompt-injected model can drive a destructive tool --- deleting a mailbox, say
--- on a server the operator never authorized at that
granularity~\cite{enclawed-site}. A lone operator on their own laptop may absorb
that risk; a bank, a hospital, or a government program putting an agent on
regulated data cannot --- and, with no admission record, cannot even prove what
happened. A
control framework such as the National Institute of Standards and Technology
(NIST) Special Publication (SP)~800-53~\cite{nist-800-53} expects
least-privilege access enforcement (AC-3, AC-6), information-flow enforcement
between sensitivity levels (AC-4), and an audit record of access decisions
(AU-2, AU-9). None of these can be satisfied if the host has no notion of which
server is which, at what level, or which tools are in bounds.

\begin{figure}[t]
\centering
\begin{minipage}[t]{0.42\linewidth}
\centering
\begin{tikzpicture}[font=\scriptsize,>={Latex},
  b/.style={draw,rounded corners=2pt,align=center,inner sep=3pt,minimum width=22mm,minimum height=7mm}]
\node[b,fill=black!5] (h) {Host\\(LLM-driven)};
\node[b,fill=black!5,below=18mm of h] (s) {MCP server};
\draw[->] ([xshift=7mm]h.south) -- node[right=0pt,font=\tiny,align=center]{\texttt{tools/}\\\texttt{call}} ([xshift=7mm]s.north);
\draw[<-] ([xshift=-7mm]h.south) -- node[left=0pt,font=\tiny,align=center]{tool\\list} ([xshift=-7mm]s.north);
\node[draw,dashed,gray,fit=(h)(s),inner sep=3.5mm,label={[font=\tiny,gray]0:TLS}] {};
\end{tikzpicture}\\[3pt]
{\scriptsize (a) \textbf{MCP today.} The host trusts the server's tool list and
dials any tool; only TLS sits between them --- no admission, sensitivity level,
or per-tool gate.}
\end{minipage}\hfill
\begin{minipage}[t]{0.54\linewidth}
\centering
\begin{tikzpicture}[font=\scriptsize,>={Latex},node distance=3mm,
  b/.style={draw,rounded corners=2pt,align=center,inner sep=3pt,minimum width=48mm,minimum height=6mm},
  g/.style={b,fill=blue!8}]
\node[b,fill=black!5] (h) {Host (LLM-driven)};
\node[g,below=of h] (v) {verify clearance assertion vs.\ the pinned\\ trust root: capability, signer, expiry,\\ level, signature, origin};
\node[g,below=of v] (a) {per-server tool allowlist};
\node[g,below=of a] (au) {append tamper-evident audit record};
\node[b,fill=black!5,below=of au] (s) {MCP server};
\draw[->] (h)--(v); \draw[->] (v)--(a); \draw[->] (a)--(au);
\draw[->] (au)--node[right=0pt,font=\tiny]{\texttt{tools/call}}(s);
\end{tikzpicture}\\[3pt]
{\scriptsize (b) \textbf{Proposed (attested admission).} The host verifies a
signed clearance assertion against a pinned trust root, enforces a closed
per-server tool allowlist, and audits the decision before any
\texttt{tools/call} --- all above MCP, no message changed.}
\end{minipage}
\caption{MCP today (a) versus the proposed attested tool-server admission (b).
The extension adds a host-side gate and one server-published document; an
unextended host ignores it and behaves as in (a).}
\label{fig:before-after}
\end{figure}
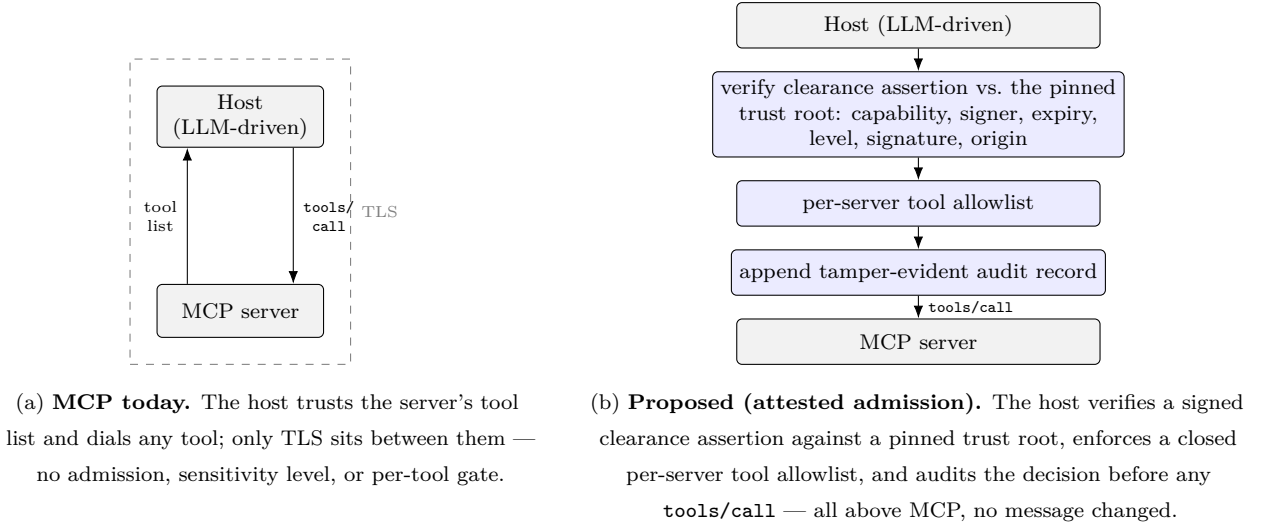

\paragraph{Contribution.}
This paper documents \texttt{mcp-attested}, the MCP admission layer, and
generalizes it into a standards-style proposal. The design rests on a single
observation: \emph{MCP needs a trust layer, but that layer can be added entirely
above the existing protocol --- one server-published document and one host-side
gate --- without changing a single existing MCP message} (\cref{fig:before-after}).
Concretely we contribute:

\begin{enumerate}[leftmargin=1.4em,itemsep=2pt]
  \item \textbf{A clearance assertion format} (\cref{sec:assertion}): a tiny,
        offline-signed JSON document, published by a server at the well-known
        URI \wellknown, that binds the server's identity to a sensitivity
        \emph{clearance} and is verifiable against a host-local trust root. We
        reuse the exact byte-canonicalization and Ed25519~\cite{rfc8032}
        signing flow already used for signing Enclawed extension manifests, so
        a server operator signs a clearance assertion with the same offline
        tool they already use for code.
  \item \textbf{An admission algorithm} (\cref{sec:admission}): the host
        verifies signature, signer authority, and clearance ordering
        \emph{before} the first tool dispatch, and treats admitting a server as
        distinct from authorizing its tools by binding each admitted server to
        a closed, per-server tool allowlist (\cref{sec:tool-allowlist}).
  \item \textbf{A flavor-gated enforcement model} (\cref{sec:flavor}): in the
        open flavor the checks are advisory (warn-but-allow, preserving
        drop-in compatibility); in the enclaved flavor they are hard denials.
        Every decision --- allow, deny, or warn --- is appended to a
        hash-chained audit log.
  \item \textbf{An adversarial evaluation} (\cref{sec:eval}): we turn each stated
        property into an executable test driven by an LLM-generated red-team
        corpus (tool-name evasions and forged clearance assertions) and report
        the results --- every evasion denied with zero leaked network writes and
        every forgery denied at the expected guard, across both a hand-written
        suite and a large model-generated coverage campaign.
  \item \textbf{A security analysis and a standardization argument}
        (\cref{sec:analysis,sec:standardization}): we state the properties the
        extension provides, the residuals it does not, its FIPS-mode
        compatibility, and why it is small and generic enough to be an MCP
        extension rather than a fork.
\end{enumerate}

We are explicit about scope. This is an \emph{admission and information-flow}
extension. It does not by itself stop a malicious admitted server from
misbehaving \emph{within} its granted tools, nor does it police the content
crossing an admitted connection; those are the jobs of, respectively, skill
capability containment~\cite{metere2026skills} and the egress reference
monitor~\cite{metere2026covert}, which compose with this layer but are out of
scope here. What this layer guarantees is that \emph{the host only ever speaks
to servers whose identity and clearance it has cryptographically checked, only
ever drives tools it has pre-authorized on those servers, and records every
such decision.}

\paragraph{Availability and status.}
This is a report on a shipping system, not a proposal. Every mechanism described
here --- the clearance assertion and its verifier, the per-server tool allowlist,
flavor-gated enforcement, and the hash-chained audit log ---
is implemented and runs in production in \emph{both} the open
\texttt{enclawed-oss} distribution and the accredited \texttt{enclawed-enclaved}
build. The only parts not yet shipped are the items \cref{sec:limits} explicitly
marks as future work (per-tool clearance, per-assertion freshness/revocation, and
submission of the normative extension).

\section{Background: MCP and its security posture}

\subsection{What MCP specifies}
MCP defines a client--server protocol carried over JSON-RPC 2.0. A
\emph{host} embeds one or more \emph{clients}; each client maintains a session
with a \emph{server} that exposes some combination of \emph{tools} (callable
functions), \emph{resources} (readable content), and \emph{prompts}. After an
\texttt{initialize} handshake that negotiates capabilities and protocol
version, the standard interaction for tool use is:
\begin{itemize}[leftmargin=1.4em,itemsep=1pt]
  \item \texttt{tools/list} $\rightarrow$ an array of tool descriptors, each
        with a name, a human-readable description, and a JSON-Schema for its
        arguments;
  \item \texttt{tools/call} with \texttt{\{ name, arguments \}}
        $\rightarrow$ a structured result.
\end{itemize}
The transport may be a local subprocess pipe (stdio) or a network channel
(HTTP, optionally with server-sent-event streaming). When the channel is
network-facing, confidentiality and server endpoint authentication are whatever
the underlying TLS~\cite{rfc8446} provides; recent revisions of the spec layer
OAuth~\cite{rfc6749} on top for \emph{user} authorization to a server.

\subsection{What MCP does not specify}
\label{sec:gap}
The protocol's trust assumptions, made explicit:
\begin{enumerate}[leftmargin=1.4em,itemsep=2pt]
  \item \textbf{No host-side server admission.} TLS authenticates that the host
        reached the endpoint named in the Uniform Resource Locator (URL); OAuth authorizes the
        \emph{user} to the server. Neither answers the host's question: ``is
        this server one I am \emph{permitted to use as a tool provider},
        and at what sensitivity?'' A valid certificate for
        \texttt{evil.example} is still a valid certificate.
  \item \textbf{No sensitivity / clearance notion.} There is no field in which
        a server declares the data classification it is cleared to handle, and
        no host-side check that a server's level dominates the level of the
        data the agent is about to send it. Information-flow control
        (Bell--LaPadula~\cite{bell-lapadula}, AC-4) has no hook.
  \item \textbf{The tool list is self-asserted and trusted wholesale.} The host
        learns the available tools from the server itself, and the typical host
        exposes all of them to the model. Admitting a server is implicitly
        admitting every tool it advertises, now and after any future
        \texttt{tools/list} change.
  \item \textbf{No standard audit of admission decisions.} Because there is no
        admission step, there is nothing to record. AU-2/AU-9 evidence must be
        reconstructed from transport logs, if at all.
\end{enumerate}
These gaps are not bugs in MCP; they are out of its chosen scope. The
contribution of this paper is to fill them \emph{above} the protocol, additively.

\subsection{The Enclawed setting}
Enclawed is a security-hardened distribution of an open agentic artificial-intelligence (AI) gateway,
shipped in two flavors. The \emph{open} flavor preserves drop-in compatibility
with the upstream catalog and prefers warnings to denials. The \emph{enclaved}
flavor is deny-by-default and enforces Federal Information Processing Standard
(FIPS)~140-3~\cite{nist-fips-140-3}
approved cryptography; it is the flavor an operator accredits against a control
framework~\cite{enclawed-site}. The same code path runs in
both; a single runtime \texttt{flavor} value selects advisory versus
enforcing behavior. \texttt{mcp-attested} is the component that brings MCP
connections under this regime, in either flavor.

\subsection{Running example: the Google Workspace bridge}
\label{sec:running-example}
We use one concrete deployment throughout --- the one that motivated the work
--- and keep it faithful to the extension's own executable test. The Enclawed
agent is to use Google's Gmail MCP server (endpoint
\texttt{gmailmcp.googleapis.com/mcp/v1}) at the \texttt{internal} clearance
level. Google operates the endpoint; Enclawed does not, and it publishes no
Enclawed clearance assertion. The bundled, signed bridge extension
\texttt{mcp-google-workspace} registers that endpoint with a \emph{closed} tool
allowlist of Gmail's read/compose tools (e.g.\ \texttt{list\_labels},
\texttt{search\_threads}, \texttt{create\_draft}). Three properties must hold
simultaneously:
(i) the connection is a deliberate, recorded admission, not an implicit
consequence of the model dialing a URL;
(ii) only the allowlisted tools are reachable --- a prompt-injected request for
a destructive tool such as \texttt{delete\_everything} is refused
\emph{before any network call}, whatever the server's \texttt{tools/list}
advertises; and
(iii) none of this changes Enclawed's own tool API, so the user's existing
skills keep working.
This is exactly the scenario the shipped admission test
(\texttt{client-admission.test.ts}) exercises, and it passes against the shipping
code. \Cref{sec:flavor} gives the connection path for servers like Google that
publish no clearance assertion of their own; \cref{sec:tool-allowlist} the tool
gate; and \cref{sec:assertion,sec:admission} the general case of a server that
\emph{does} publish one, of which the Google bridge is the
no-server-side-assertion special case.

\section{Threat model}
\label{sec:threat}

\paragraph{Deployment.}
We assume the Enclawed personal-assistant trust model: one operator (or one
operating organization) per gateway instance; the host and its locally pinned
trust root are part of the trusted computing base; the LLM is \emph{not}.
The agent's instruction stream is assumed to be partially attacker-controlled
through retrieved content, tool outputs, and message bodies --- the standard
prompt-injection posture~\cite{owasp-llm-top-10,mitre-atlas}.

\paragraph{Adversary capabilities.}
We consider an adversary who can:
\begin{enumerate}[leftmargin=1.4em,itemsep=1pt]
  \item inject instructions into the model's context that steer it toward
        connecting to an attacker-chosen MCP server or invoking an
        attacker-chosen tool;
  \item stand up a network MCP server with a valid TLS certificate for a domain
        they control;
  \item present a self-declared tool list and a self-declared sensitivity
        level of their choosing;
  \item replay or tamper with a legitimately-issued clearance assertion in
        transit (the host--server channel is assumed adversary-observable
        unless protected by TLS).
\end{enumerate}

\paragraph{Adversary limits.}
The adversary cannot forge an Ed25519 signature without a trust-root private
key; cannot mutate the pinned trust root, which is locked after startup
(\cref{sec:trustroot}); and is not, in the base model, a co-tenant on the host
--- the host and its pinned trust root are part of the trusted computing base.
Multi-tenant adversarial isolation on one shared gateway is explicitly out of
scope, consistent with the framework's stated security model.

\paragraph{Goals.} Against this adversary the extension must guarantee:
(G1) the host never \emph{enforces a connection as admitted} to a server whose
clearance assertion is missing, malformed, unsigned, signed by a key not in the
trust root, signed by an expired key, signed by a key not authorized for the
asserted level, whose asserted level is below what the call requires, or that is
presented from a host outside the assertion's signed origin list;
(G2) the host never dispatches a tool the operator did not pre-authorize for
that server, even if the server advertises it and the model requests it;
(G3) every admission decision is recorded in tamper-evident storage; and
(G4) the pinned trust root cannot be mutated at runtime to admit an
otherwise-failing server.

\paragraph{Empirically demonstrated runtime exploits.}
The trust gap above is not theoretical. Adversarial testing against unprotected
multi-server MCP deployments has surfaced three exploitable runtime behaviors
that bypass standard post-invocation defenses:
\begin{enumerate}[leftmargin=1.4em,itemsep=1pt]
  \item \textbf{Tool-name shadowing.} An untrusted or injected server registers
        a tool name that collides with a co-loaded legitimate tool surface,
        causing the host to route execution calls silently to the malicious
        server.
  \item \textbf{Dynamic schema drift.} A malicious server advertises a benign
        schema during the initial \texttt{tools/list} handshake to pass static
        filters, then mutates parameter constraints or semantic requirements
        before the subsequent \texttt{tools/call} invocation.
  \item \textbf{Registry poisoning.} Direct manipulation of the host's tool
        cache via downstream prompt injection or unvetted secondary discovery
        layers.
\end{enumerate}
Defender-side benchmarks across these
vectors~\cite{algovoi-agent-trust-bench} show that relying on the model
layer to detect them is a brittle,
non-deterministic boundary: permissive personas completely fail to identify
active schema drift, while catching tool-name shadowing imposes a reasoning
load that degrades rapidly under context-window exhaustion. ATSA's contribution is to move the security
boundary from probabilistic, post-invocation reasoning to deterministic,
cryptographic admission-time gating --- preventing the prompt-injected model
from ever accessing or driving an unvetted tool surface in the first place.
Vectors 1 and 3 are closed at admission and tool-authorization time
(\cref{sec:atsa}); vector 2 --- schema drift after a server is admitted ---
is structurally adjacent and is the subject of
\cref{sec:drift-composition}.

\section{The clearance assertion}
\label{sec:assertion}

The first new artifact is a server-published \emph{clearance assertion}: a
small JSON document, served at the well-known URI~\cite{rfc8615}
\wellknown, that binds a server's identity to a declared sensitivity level and
is signed offline by a key the host trusts. Its shape is deliberately identical
to the Enclawed extension manifest (\texttt{enclawed.module.json}), so a server
operator signs it with the same offline procedure they already use to sign
extension code, and the host verifies it with the same routines.

\begin{lstlisting}[language=,caption={A clearance assertion served at \texttt{/.well-known/enclawed-clearance.json}.},captionpos=b]
GET https://gmail-mcp.internal.example/.well-known/enclawed-clearance.json
->  {
      "v": 1,
      "id":           "mcp.example.gmail",
      "publisher":    "example-corp",
      "version":      "2.3.1",
      "clearance":    "restricted-plus",   // generic ladder;
                                           // US-gov alias e.g. "q-cleared"
      "capabilities": ["mcp-server"],
      "signerKeyId":  "example-corp-prod-2026",
      "signature":    "<base64 Ed25519 over the canonical body>"
    }
\end{lstlisting}

\paragraph{Fields.}
\texttt{clearance} is a level in an ordered classification lattice
(\cref{sec:lattice}). \texttt{capabilities} must contain the literal
\texttt{"mcp-server"}; this is what distinguishes a clearance assertion for an
MCP endpoint from any other signed manifest using the same envelope.
\texttt{signerKeyId} names a key the host expects to find in its trust root,
and \texttt{signature} is an Ed25519 signature over the \emph{canonical
serialization} of every field except the signature itself. Two optional fields
complete the envelope and also fall under the signature: \texttt{netAllowedHosts}
binds the assertion to specific origins (enforced in \cref{sec:admission}), and
\texttt{verification} records the producer's verification level.

\begin{definition}[Canonical body]
\label{def:canonical}
The canonical body of an assertion comprises exactly the fields \texttt{v},
\texttt{id}, \texttt{publisher}, \texttt{version}, \texttt{clearance},
\texttt{capabilities}, \texttt{signerKeyId}, \texttt{verification}, and
\texttt{netAllowedHosts}; the \texttt{signature} field is excluded, and an absent
\texttt{signerKeyId} is serialized as \texttt{null}. It is emitted as
deterministic JSON in which every object's keys are written in sorted
(lexicographic) order and array members are sorted. Both signer and verifier
compute exactly these bytes, so signing and verification cannot disagree on
field order or whitespace.
\end{definition}

Using one canonicalization for extension manifests \emph{and} MCP clearance
assertions is a deliberate attack-surface reduction: there is a single signing
path, a single verification path, and a single place where a canonicalization
bug could live, rather than one per document type.

\subsection{The classification lattice}
\label{sec:lattice}

Clearance levels are not free-form strings; they are points in a totally
ordered lattice with a numeric rank, so ``does server level dominate required
level'' is a single integer comparison. The framework ships five built-in
schemes --- \texttt{default}, \texttt{us-government}, \texttt{healthcare-hipaa},
\texttt{financial-services}, and a \texttt{generic-3-tier} --- and validates
custom schemes loaded from JSON (ranks must be contiguous from~$0$). Three of
the built-ins:

\begin{center}\small
\begin{tabular}{clll}
\toprule
Rank & \texttt{default} & \texttt{us-government} & \texttt{healthcare-hipaa} \\
\midrule
0 & \texttt{PUBLIC}         & \texttt{UNCLASSIFIED} & \texttt{PUBLIC} \\
1 & \texttt{INTERNAL}       & \texttt{CUI}          & \texttt{INTERNAL} \\
2 & \texttt{CONFIDENTIAL}   & \texttt{CONFIDENTIAL} & \texttt{PHI} \\
3 & \texttt{RESTRICTED}     & \texttt{SECRET}       & \texttt{SENSITIVE-PHI} \\
4 & \texttt{RESTRICTED-PLUS}& \texttt{TOP SECRET}   & \texttt{RESEARCH-EMBARGOED} \\
5 & \texttt{SCI}            & \texttt{TS//SCI}      & --- \\
\bottomrule
\end{tabular}
\end{center}

Each level carries aliases so an organization can speak its own dialect
(\texttt{SECRET} for \texttt{RESTRICTED}, \texttt{CUI} for \texttt{INTERNAL})
while the engine compares ranks; the regulated schemes additionally define valid
compartments (\texttt{MENTAL-HEALTH}, \texttt{GENETICS}, \texttt{M\_AND\_A},
\dots) and releasability markings (\texttt{NOFORN}, \texttt{BAA-COVERED},
\dots). This is what makes the same admission code sector-neutral: a
\texttt{financial-services} deployment ranks \texttt{MNPI} above
\texttt{CONFIDENTIAL}, a hospital ranks \texttt{SENSITIVE-PHI} above
\texttt{PHI}, a government program uses the US-government ladder --- the
verifier is identical. (Markings above: \texttt{CUI}, Controlled Unclassified
Information; \texttt{SCI}, Sensitive Compartmented Information; \texttt{PHI},
Protected Health Information; \texttt{MNPI}, Material Non-Public Information;
\texttt{NOFORN}, Not Releasable to Foreign Nationals; \texttt{BAA}, Business
Associate Agreement.)

\begin{definition}[Domination]
\label{def:dominates}
Let $\mathrm{rank}(\ell)$ be the numeric rank of level $\ell$ in the applicable
ladder (level names are matched case-insensitively, canonical name or alias). A
server at level $s$ \emph{meets} a required level $r$, written $s \succeq r$,
iff $\mathrm{rank}(s) \ge \mathrm{rank}(r)$.
\end{definition}

\section{The admission algorithm}
\label{sec:admission}

When the host is about to use an MCP server, it runs the verification below
\emph{before} treating the connection as admitted. The fetch step is injectable
(so it can run over the mutually-authenticated TLS channel inside the enclave
network, or be stubbed in tests); everything after it is pure and
side-effect-free except for the audit append.

\begin{lstlisting}[language=,caption={Server clearance verification (\texttt{verifyServerClearance}), faithful to the shipping implementation.},captionpos=b]
verify(serverUrl, required):
  doc  <- GET serverUrl + "/.well-known/enclawed-clearance.json"
          | on HTTP error or network failure: DENY("fetch failed")
  m    <- parseManifest(doc)
          | on malformed: DENY("manifest invalid")
  if "mcp-server" not in m.capabilities:        DENY("not an mcp-server manifest")
  if m.signerKeyId absent or m.signature absent: DENY("unsigned")
  signer <- trustRoot.find(m.signerKeyId)
  if signer is null:                            DENY("signer not in trust root")
  if signer.notAfter and signer.notAfter < now: DENY("signer expired")
  if m.clearance not in signer.approvedClearance: DENY("signer not approved for level")
  if not ed25519_verify(canonicalBody(m), m.signature, signer.publicKey):
                                                DENY("signature verification failed")
  if not meets(m.clearance, required):          DENY("server level below required")
  if m.netAllowedHosts nonempty and host(serverUrl) not in m.netAllowedHosts:
                                                DENY("not valid for host")  // origin binding
  ALLOW(clearance = m.clearance, signerKeyId = signer.keyId)
\end{lstlisting}

The order is security-relevant. Capability and signedness are cheap structural
checks; signer lookup and the \texttt{approvedClearance} check establish
\emph{authority} (does the trust root let \emph{this} signer vouch for
\emph{this} level?) \emph{before} the cryptographic verification; the signature
check establishes \emph{authenticity}; and only then does the level comparison
establish \emph{sufficiency}. A server cannot self-promote: even a validly
signed assertion is denied if the signing key is not authorized in the trust
root for the level it claims.

\subsection{The trust root}
\label{sec:trustroot}

The trust root is a set of records, each pairing a \texttt{keyId} and an
Ed25519 public key with an \texttt{approvedClearance} list --- the maximum set
of levels that key is allowed to vouch for. A development bundle ships three
signers: a community signer and a bundled-extension signer, each approved only
up to \texttt{INTERNAL}, and a high-tier reference signer approved up through
\texttt{RESTRICTED-PLUS} for exactly the attested-MCP use case. Production
deployments replace these with the operator's own signers.

The crucial property is that the trust root is \emph{lockable}. The enclaved
bootstrap calls \texttt{lockTrustRoot()} immediately after seeding it; once
locked, \texttt{setTrustRoot()} refuses further mutation. This binds the running
binary to the deploying organization's signers and removes ``swap the trust
root at runtime'' from the adversary's toolkit.

\subsection{Connection gating and flavor semantics}
\label{sec:flavor}

The verifier returns a verdict; what the host \emph{does} with a negative
verdict depends on the flavor.

\begin{itemize}[leftmargin=1.4em,itemsep=2pt]
  \item \textbf{Enclaved flavor:} a negative verdict is a hard denial. The
        connection is not admitted, \texttt{invoke} returns an error, and an
        \texttt{mcp.connect.deny} record is appended to the audit log. This is
        the accreditable mode and satisfies goal G1.
  \item \textbf{Open flavor:} the \emph{same check runs}, but a negative
        verdict is surfaced as a warning rather than masked. The client returns
        a non-OK result tagged ``open flavor: warn-only'' and writes an
        \texttt{mcp.connect.warn} record, leaving the host free to proceed if it
        explicitly chooses. Critically, the open path never silently rewrites a
        failure into a success --- it refuses to return \texttt{ok} on a failed
        verification, so a warning can never be mistaken for a pass.
\end{itemize}

A successful verification in either flavor appends \texttt{mcp.connect.allow}
with the resolved clearance and signer key id. There is also a deliberate
first-party fast path: bundled bridges (e.g.\ a Google Workspace or GitHub
bridge) whose \emph{own} signed extension manifest and registry registration
already constitute the admission decision may set
\texttt{skipClearancePreflight}; this path still audits the allow, so no
admitted connection is unrecorded. The fast path exists because public
third-party MCP servers do not (yet) publish an Enclawed clearance assertion,
and the bridge's signed manifest is the equivalent gate
(\cref{sec:standardization} returns to this).

\subsection{Tool-level admission}
\label{sec:tool-allowlist}

Admitting a \emph{server} is not authorizing its \emph{tools}. This is the
second new gate and the answer to gap~3 of \cref{sec:gap}. Each server the host
will use is recorded in a registry entry binding the endpoint to: the owning
bridge, the required clearance, a \emph{closed} \texttt{allowedTools} list, and
the transport. On every \texttt{invoke}:

\begin{lstlisting}[language=,caption={Tool dispatch with per-server allowlist enforcement.},captionpos=b]
invoke(serverUrl, toolName, args):
  entry <- registry.byEndpoint(serverUrl)
  if entry exists:
    if toolName not in entry.allowedTools:
      audit("mcp.tool.deny", {server, bridge, toolName, "not in allowedTools"})
      return DENY
    if not connect(serverUrl).ok:    return DENY      // re-runs clearance gate
    return entry.transport.call("tools/call", {name: toolName, arguments: args})
  else:
    connect(serverUrl)               // gate still runs
    return DENY("no registered bridge; register the endpoint explicitly")
\end{lstlisting}

Three things follow. First, the allowlist is a \emph{closed} list: a tool the
model requests that is not on it is denied and audited (\texttt{mcp.tool.deny}),
regardless of what \texttt{tools/list} advertised --- this is goal G2. Second,
the clearance gate is re-run at dispatch, not just at first connect, so a
connection cannot be admitted once and then reused indefinitely without
re-checking. Third, an endpoint with no registry entry cannot be driven at all:
there is no implicit ``trust whatever the model dialed'' path. The combination
means a prompt-injected model that asks to call
\texttt{gmail.deleteAllMessages} on a server whose allowlist contains only
\texttt{gmail.search} is denied at the host, before any byte reaches the
network.

\subsection{Transport}
Dispatch rides the existing MCP envelope unchanged: a JSON-RPC 2.0
\texttt{tools/call} over HTTP(S), with an optional \texttt{Authorization:
Bearer} token supplied by a pluggable auth provider for servers that also do
OAuth user-authorization. The extension adds \emph{no} new message types to
MCP; the clearance assertion is fetched out-of-band at a well-known URI, and
the tool allowlist is enforced entirely host-side. An unextended MCP server is
unaware that any of this happened.

\subsection{Audit events}
\label{sec:audit}

Every decision is an append to the framework's hash-chained, optionally
multi-witnessed audit log~\cite{enclawed-site}. The event vocabulary is
small and total over the decision space:

\begin{center}\small
\begin{tabular}{lll}
\toprule
Event & When & Carries \\
\midrule
\texttt{mcp.connect.allow} & verification passed (or bridge fast-path) & level, signerKeyId \\
\texttt{mcp.connect.deny}  & verification failed, enclaved flavor      & reason \\
\texttt{mcp.connect.warn}  & verification failed, open flavor          & reason \\
\texttt{mcp.tool.deny}     & tool not in server allowlist              & bridge, toolName \\
\bottomrule
\end{tabular}
\end{center}

Because the log is hash-chained, the sequence of admission decisions is
tamper-evident: an auditor can replay it and detect any excision or
back-dating, which is what AU-9 asks for and what goal G3 requires.

\section{Security analysis}
\label{sec:analysis}

We restate the guarantees as properties and argue each from the mechanism.

\begin{property}[Authenticated, authorized admission]
\label{prop:g1}
In the enclaved flavor, a connection is enforced as admitted only if the
server's clearance assertion (a) declares the \texttt{mcp-server} capability,
(b) is signed, (c) by a key present in the locked trust root, (d) whose
\texttt{notAfter} (if set) has not passed, (e) that is approved for the asserted
level, (f) the signature verifies over the canonical body, (g) the asserted
level dominates the required level, and (h) when the assertion carries a signed
host allow-list, the connected host is a member of it.
\end{property}
\noindent\emph{Argument.} Each clause is a guarded \texttt{DENY} in
\cref{sec:admission} that precedes \texttt{ALLOW}; the conjunction is necessary.
Forging (f) requires a trust-root private key (adversary limit); bypassing (c),
(d), or (e) requires mutating the trust root, which is locked after startup
(\cref{sec:trustroot}). A validly-signed assertion cannot have its level
escalated, because the level is inside the signed body. Clause~(h) closes
cross-host replay: because \texttt{netAllowedHosts} is part of the signed
canonical body, an attacker cannot retarget a valid assertion to a host the
signer did not authorize without invalidating the signature; an assertion that
omits the list remains origin-agnostic and relies, as before, on the enclave's
mutually-authenticated TLS and trust-root scoping. Our adversarial harness
(\cref{sec:eval}) confirms clauses (a)--(h) fire as guarded denials, including a
host-bound assertion replayed from a foreign origin and an assertion signed by
an expired signer. $\qed$

\begin{property}[Tool least-privilege]
\label{prop:g2}
An admitted server's tool is dispatched only if it appears in that server's
closed \texttt{allowedTools} list, independent of what the server advertises or
the model requests.
\end{property}
\noindent\emph{Argument.} \texttt{invoke} checks \texttt{isToolAdmitted} against
the registry entry before any transport call and denies otherwise; an
unregistered endpoint has no dispatch path at all (\cref{sec:tool-allowlist}).
The server's \texttt{tools/list} feeds the model's \emph{options} but not the
host's \emph{authorization}. $\qed$

\begin{property}[Decision auditability]
Every admission and tool-gating decision appends one event to a hash-chained
log; the event set is total over \{allow, deny, warn, tool-deny\}.
\end{property}
\noindent\emph{Argument.} Each return path in \texttt{connect} and
\texttt{invoke} is preceded by an audit append (\cref{sec:audit}); the chain is
tamper-evident by construction. $\qed$

\begin{property}[Trust-root immutability]
Once locked at startup, the pinned trust root cannot be mutated, so the set of
signers under which a server may be admitted cannot be widened at runtime.
\end{property}
\noindent\emph{Argument.} \texttt{lockTrustRoot()} sets a one-way flag;
thereafter \texttt{setTrustRoot()} raises \texttt{TrustRootLockedError}
(\cref{sec:trustroot}). Integrity of the host binary itself is a
trusted-computing-base assumption (\cref{sec:threat}), not a property this layer
claims. $\qed$

\paragraph{Residuals (explicitly out of scope).}
The extension does \emph{not} provide: (i) protection against a \emph{correctly
admitted, correctly cleared} server that misbehaves within its allowed tools ---
that is the job of capability containment on the skill/tool
side~\cite{metere2026skills}; (ii) content inspection of data crossing an
admitted connection --- that is the egress reference
monitor~\cite{metere2026covert} and the data-loss-prevention (DLP) scanner, which run downstream;
(iii) defense against an LLM that simply refuses to use a tool (a liveness, not
safety, concern); (iv) multi-tenant adversarial isolation on one shared gateway,
which the framework's trust model excludes by design. These layers compose: a
call must pass admission \emph{and} clearance \emph{and} tool-allowlist
(this paper) \emph{and} capability containment \emph{and} egress screening
before it leaves the host.

\paragraph{FIPS posture.}
The admission path uses only Ed25519 (FIPS~186-5) and the Secure Hash Algorithm
SHA-256 (FIPS~180-4), routed through the framework's approved-algorithm
enforcement, so it runs unmodified in a FIPS-mode enclaved deployment.

\section{Evaluation}
\label{sec:eval}

The properties of \cref{sec:analysis} are stated as arguments from the
mechanism; this section turns each into an executable test and reports the
result. The tests ship in the open \texttt{enclawed-oss} reference
implementation (\texttt{extensions/mcp-attested}) and are reproducible with
\texttt{npx vitest run \dots/adversarial.test.ts}.

\subsection{Methodology}

\paragraph{Hermetic, network-observing harness.}
Every test runs against in-memory stubs, never a live server. Two seams make
this faithful rather than merely convenient: the verifier's \texttt{fetcher}
argument supplies the clearance assertion, and the transport's
\texttt{fetchImpl} supplies tool-call responses. The transport stub increments a
counter on every call, so a denial that nonetheless leaked a network write would
be caught by an assertion that the counter is \emph{zero}. A fake runtime
installed through \texttt{setRuntime()} records every audit append, so the audit
vocabulary of \cref{sec:audit} is checked directly rather than inferred.

\paragraph{Faithful positives, single-mutation negatives.}
Trust roots are seeded per test with freshly generated Ed25519 keypairs. Valid
baselines are signed through exactly the production path
(\texttt{canonicalManifestBytes} $\to$ \texttt{signManifest}); each negative is
one targeted mutation of a baseline --- drop the signature, swap the signer,
upgrade the \texttt{clearance} field \emph{after} signing, omit the
\texttt{mcp-server} capability, and so on --- so a test isolates exactly one
guard of \cref{prop:g1}.

\paragraph{LLM-generated adversarial corpus.}
The inputs are a red-team corpus authored by the LLM as if it were the attacker.
For the tool-name gate it spans command-chaining (\texttt{list\_labels;
delete\_everything}), case folding, whitespace and control-character injection,
Unicode homoglyph / right-to-left-override (RTL) / zero-width tricks, path traversal,
near-miss truncations, prototype-pollution probe strings (\texttt{\_\_proto\_\_},
\texttt{constructor}), and JSON smuggling. For the admission gate it spans
unsigned, wrong-signer, signer self-promotion, signature tamper, post-signing
body upgrade, capability omission, level shortfall, and malformed / transport
error assertions.

\subsection{Results}

All \textbf{48} tests across the \texttt{mcp-attested} admission layer and its
Google Workspace bridge pass (44 under \texttt{vitest}, 4 under
\texttt{node:test}); the 20 adversarial tests are summarized below.

\begin{center}\small
\begin{tabular}{p{0.31\linewidth}cp{0.40\linewidth}}
\toprule
Property under test & Inputs & Outcome \\
\midrule
\cref{prop:g2} tool least-privilege & 31 evasions & all 31 denied (\texttt{mcp.tool.deny}); \textbf{0} network writes; only the 2 exact allowlisted names admitted \\
\addlinespace
\cref{prop:g1} authenticated, authorized admission & 2 valid + 12 invalid & both valid admitted (incl.\ a host-bound assertion from its own origin); all 12 denied, each at the expected guard (unsigned, not-in-root, expired-signer, not-approved, bad-signature, below-required, host-not-bound, \dots) \\
\addlinespace
Auditability + flavor gating & 3 paths & enclaved fail $\to$ \texttt{mcp.connect.deny}; open fail $\to$ \texttt{mcp.connect.warn} and never \texttt{ok}; success $\to$ \texttt{mcp.connect.allow} \\
\bottomrule
\end{tabular}
\end{center}

\paragraph{The harness is adversarial in substance.}
Beyond confirming intended denials, the corpus surfaced a genuine origin-binding
weakness in an earlier verifier --- which is why the verifier now enforces the
signed host allow-list (\cref{prop:g1}(h)). A from-scratch adversarial corpus
that forces a real fix, rather than only re-confirming the happy path, is the
evidence that the suite earns its name.

\subsection{Dynamic and live conformance (gated)}

Two further harnesses ship alongside the hermetic suite but depend on external
prerequisites, so we describe their method rather than report numbers.

\paragraph{LLM-driven fuzzing.}
A local model served by Ollama is asked, at run time, to red-team the gate ---
to emit tool-name evasions against a bridge allowlist and to forge clearance
assertions that might slip past the verifier. Each generated tool name that is
not an exact allowlist member must be denied with no network write, and each
forged assertion must be rejected. Crucially, the security invariant is
independent of the model's quality: the model never holds a trust-root private
key, so it cannot produce a verifying signature, and exact-match admission is
immune to its surface tricks --- a stronger model merely yields more inventive
denials. The harness is enabled with \texttt{OLLAMA\_LIVE=1} and fails loudly if
no model is reachable, so it never passes vacuously.

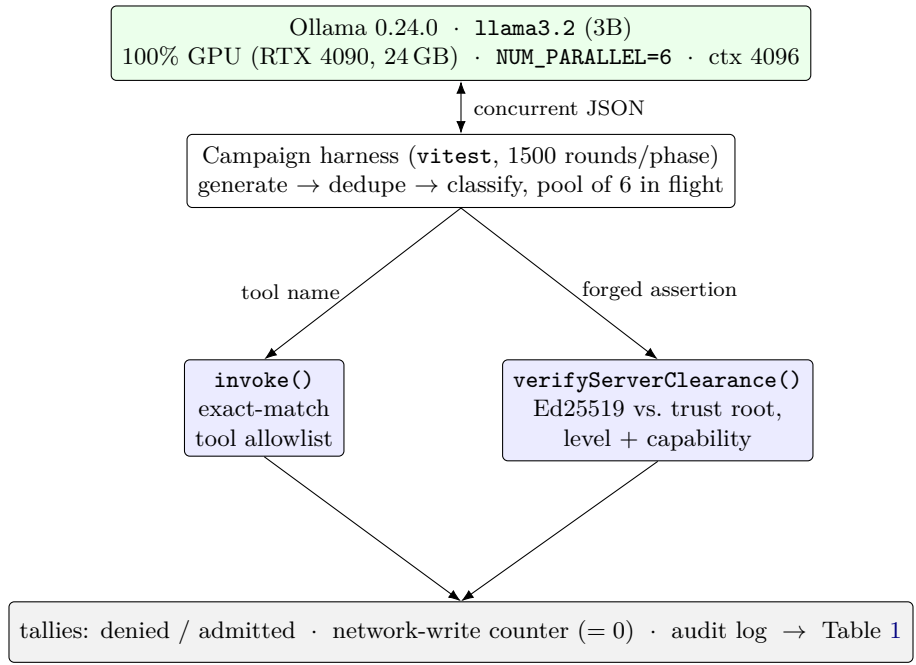
\begin{figure}[h]
\centering
\begin{tikzpicture}[
  font=\footnotesize, >={Latex}, node distance=7mm and 6mm,
  box/.style={draw, rounded corners=2pt, align=center, inner sep=4pt, minimum height=8mm},
  gen/.style={box, fill=green!8},
  gate/.style={box, fill=blue!8},
  sink/.style={box, fill=black!5},
]
\node[gen] (ollama) {Ollama 0.24.0 $\;\cdot\;$ \texttt{llama3.2} (3B)\\
  100\% GPU (RTX 4090, 24\,GB) $\;\cdot\;$ \texttt{NUM\_PARALLEL=6} $\;\cdot\;$ ctx 4096};
\node[box, below=of ollama] (harness) {Campaign harness (\texttt{vitest}, 1500 rounds/phase)\\
  generate $\to$ dedupe $\to$ classify, pool of 6 in flight};
\node[gate, below=20mm of harness, xshift=-26mm] (invoke) {\texttt{invoke()}\\
  exact-match\\ tool allowlist};
\node[gate, below=20mm of harness, xshift=26mm] (verify) {\texttt{verifyServerClearance()}\\
  Ed25519 vs.\ trust root,\\ level + capability};
\node[sink, below=52mm of harness] (out) {tallies: denied / admitted $\;\cdot\;$ network-write
  counter ($=0$) $\;\cdot\;$ audit log $\;\to\;$ \cref{tab:campaign}};
\draw[<->] (ollama) -- node[right=1pt,font=\scriptsize]{concurrent JSON} (harness);
\draw[->] (harness.south) -- node[left=1pt,pos=0.55,font=\scriptsize]{tool name} (invoke.north);
\draw[->] (harness.south) -- node[right=1pt,pos=0.55,font=\scriptsize]{forged assertion} (verify.north);
\draw[->] (invoke.south) -- (out.north);
\draw[->] (verify.south) -- (out.north);
\end{tikzpicture}
\caption{Runtime configuration of the adversarial campaign. A GPU-resident local
model (Ollama / \texttt{llama3.2}) generates adversarial tool names and forged
clearance assertions; the harness deduplicates and feeds them through the two
deterministic gate paths; outcomes (denials, admitted, network writes) are
tallied into \cref{tab:campaign}.}
\label{fig:runtime}
\end{figure}

We ran a coverage campaign with \texttt{llama3.2} served by Ollama~0.24.0
(\cref{fig:runtime}: 1500 generation rounds per phase at 6-way graphics-processing-unit (GPU) concurrency,
deduplicated). Because the gate is deterministic
--- exact-match for tool names, an Ed25519 signature for assertions --- a
passing result is full \emph{coverage}, not a deny probability:
\cref{tab:campaign} reports how many distinct adversarial inputs the model
produced and how many the gate denied. Of \textbf{27{,}025} unique tool-name
evasions (from 47{,}945 generated) spanning whitespace/control,
separator-chaining, near-miss, path-traversal, homoglyph/zero-width/RTL, and
case-variant tricks, \emph{all} were denied with \emph{zero} network writes and
none admitted. Of the model's forged clearance assertions, \textbf{14{,}378}
distinct ones survived JSON parsing and \emph{all} were rejected (the model
holds no trust-root private key, so a verifying signature is out of reach;
malformed forgeries are discarded by the harness and would be denied in any
case). This attacker-side campaign is complemented by independent defender-side
baselines from the AlgoVoi \emph{Agent-Trust-Bench}
suite~\cite{algovoi-agent-trust-bench}, whose differential profiles across
29-tool surfaces confirm that unprotected deployments remain uniformly
vulnerable to the runtime routing and session exploits enumerated in
\cref{sec:threat}.

\begin{table}[h]
\centering\small
\caption{LLM-driven coverage campaign (\texttt{llama3.2}, 1500 rounds/phase):
every distinct adversarial input the model generated was denied by the gate.}
\label{tab:campaign}
\begin{tabular}{lrrr}
\toprule
Tool-name evasion category & unique & denied & admitted \\
\midrule
unclassified / other          & 12{,}106 & 12{,}106 & 0 \\
whitespace / control char     &  9{,}012 &  9{,}012 & 0 \\
separator / command chaining  &  3{,}469 &  3{,}469 & 0 \\
near-miss (edit dist.\ $\le$2) &  1{,}251 &  1{,}251 & 0 \\
path-traversal                &    752 &    752 & 0 \\
homoglyph / zero-width / RTL   &    393 &    393 & 0 \\
case-variant of an allow name  &     42 &     42 & 0 \\
\midrule
\textbf{Total} (47{,}945 generated) & \textbf{27{,}025} & \textbf{27{,}025} & \textbf{0} \\
\midrule
\multicolumn{4}{l}{\emph{Network writes on denied calls:} 0\qquad
  \emph{Forged assertions:} 14{,}378 unique, 14{,}378 denied, 0 admitted} \\
\bottomrule
\end{tabular}
\end{table}

\paragraph{Live end-to-end.}
A second harness uses the local model as the agent brain driving the real Google
Workspace bridge (\cref{sec:running-example}) through the gate, with
operator-supplied OAuth read only from the environment. It asserts the
deterministic invariant --- a tool outside the allowlist is denied before any
network call --- and, best-effort, that an allowlisted tool the model selects is
dispatched to the live endpoint, with any failure on that path required to be a
transport/scope error rather than an admission denial. It is gated behind
\texttt{GOOGLE\_LIVE=1} plus operator credentials. We executed it against
Google's live Gmail MCP endpoint (\texttt{gmailmcp.googleapis.com/mcp/v1},
independently confirmed reachable with an \texttt{HTTP 200} to \texttt{tools/list})
using an operator-supplied OAuth token: the gate admitted the allowlisted tool
the model selected and dispatched it to the real endpoint with no admission
denial, and denied the out-of-allowlist tool before any network call.

This keeps the headline numbers honest: the 48 hermetic tests reproduce anywhere
with no setup, while the gated LLM-driven and live tests (the Ollama fuzz, the
coverage campaign, and the live Gmail end-to-end) --- which additionally require
a local model and, for the Gmail path, operator credentials --- were run and also
pass. They are provided as operator-runnable conformance checks.

\section{Why a protocol extension, and how to standardize it}
\label{sec:standardization}

The design choices above were made so that the mechanism is a candidate for a
\emph{generic} MCP extension, not just an Enclawed feature.

\paragraph{If it already works above MCP, why change the standard?}
The natural objection: the extension is purely additive (below), so an integrator
can bolt it on today without the maintainers lifting a finger --- why ask them to
do anything? Because \emph{additive} is not \emph{interoperable}. Bolted on by one
vendor, attestation secures only that vendor's island: a server's signed clearance
document means something only to hosts that already agreed, out of band, on its
shape and verification order. The payoff --- ``this server is attested'' as a
claim \emph{any} host can check, servers publishing \emph{one} document instead of
$N$ vendor dialects, and the protocol's own SDKs shipping the gate on
\emph{by default} so security stops being each integrator's opt-in --- exists only
once a single format is agreed. And only the specification owner can mint that
agreement: no vendor can unilaterally make a well-known name and a verification
order the ecosystem's Schelling point, which is precisely why public MCP servers
publish no attestation today (\cref{sec:flavor}). The maintainers' work is small,
and is the part only they can do: reserve the well-known name and bless one
document shape and verification order (\cref{sec:atsa}) --- the reference
implementation and conformance vectors (\cref{sec:atsa-vectors}) already exist.
That the mechanism rides entirely above MCP is what makes adoption \emph{cheap and
low-risk}, not a reason to defer it: it is an optional, additive addendum with no
compatibility tax.

\paragraph{It is purely additive.} The clearance assertion lives at a
well-known URI~\cite{rfc8615}; an unextended host never fetches it and is
unaffected. The tool allowlist is host-side state; an unextended server never
sees it. No existing MCP message changes shape. This is the same additive
discipline that let OAuth, \texttt{security.txt}, and OpenID discovery ride the
well-known mechanism without forking the protocols they extended.

\paragraph{It separates the three questions MCP conflates.} TLS answers ``did I
reach the named endpoint.'' OAuth answers ``is the user allowed to use this
server.'' The clearance assertion answers the missing third question: ``is this
server one my organization has authorized as a tool provider, and at what
sensitivity.'' These are orthogonal and should be layered, not merged; a
standardized clearance assertion could sit alongside MCP's existing OAuth
profile rather than competing with it.

\subsection{Normative specification: Attested Tool-Server Admission (ATSA)}
\label{sec:atsa}

We state the extension in normative form so that it can be adopted as an MCP
addendum without further design work --- and so that no reviewer can dismiss it
as under-specified. The key words \textsc{must}, \textsc{must not},
\textsc{required}, \textsc{shall}, \textsc{should}, \textsc{should not},
\textsc{may}, and \textsc{optional} are to be interpreted as in RFC~2119 and
RFC~8174. ``Host'' and ``server'' carry their MCP meanings.
The extension adds no new MCP message type and alters no existing one
(\textsc{compatibility}, below).

\paragraph{Discovery.}
A server \textsc{may} advertise an attestation two interoperable ways. (i)
\emph{Well-known resource}: the server \textsc{should} serve a Server
Attestation Document (SAD) at \texttt{/.well-known/mcp-attestation} (the Enclawed
profile uses \wellknown) on the same origin as its MCP endpoint, over TLS. (ii)
\emph{In-band}: a server \textsc{may} return the SAD in the \texttt{initialize}
result under an experimental capability key. A conforming host \textsc{must}
attempt~(i), \textsc{may} attempt~(ii), and \textsc{must} treat the absence of
both as ``unattested,'' applying its configured posture (see
\textsc{compatibility}).

\paragraph{Server Attestation Document.}
The SAD is the vendor-neutral name for the \emph{clearance assertion} of
\cref{sec:assertion}: a JSON object whose \textsc{required} fields are:
\texttt{v} (integer \texttt{1}); \texttt{id} (stable server identity);
\texttt{publisher}; \texttt{version}; \texttt{clearance} (a canonical name or
alias of a level in a named, totally ordered classification scheme,
\cref{sec:lattice}); \texttt{capabilities} (string array that \textsc{must}
contain \texttt{"mcp-server"}); \texttt{signerKeyId} (identifies a key in the
host trust root); and \texttt{signature} (base64 detached signature over the
canonical body). Two fields are \textsc{optional} and, when present, are part of
the signed body: \texttt{netAllowedHosts} (string array) is \textsc{recommended}
--- when non-empty it binds the SAD to those origins --- and \texttt{verification}
(string, e.g.\ \texttt{tested}) records the producer's verification level. The
formal schema is \cref{sec:atsa-schema}. Verifiers \textsc{must} ignore unknown
fields and \textsc{must not} include them in the canonical body unless a future
\texttt{v} registers them.

\paragraph{Canonicalization and signature.}
The canonical body is the deterministic JSON serialization of every registered
field except \texttt{signature}, object keys in sorted order, array members
sorted, an absent \texttt{signerKeyId} serialized as \texttt{null}
(\cref{def:canonical}). Signer and verifier \textsc{must} compute the signature
over exactly these bytes. The Enclawed profile uses Ed25519~\cite{rfc8032}; a
profile \textsc{may} substitute another suite if signer and verifier agree.

\paragraph{Host verification rules.}
Before treating a connection as \emph{admitted}, a host \textsc{must} evaluate
the following in order and \textsc{must} deny on the first failure:
\begin{enumerate}[leftmargin=2.2em,itemsep=1pt,label=(\alph*)]
  \item the SAD parses and its \texttt{capabilities} contain \texttt{"mcp-server"};
  \item \texttt{signerKeyId} and \texttt{signature} are present;
  \item \texttt{signerKeyId} resolves to a key in the trust root;
  \item that key's validity window (\texttt{notAfter}, if set) includes now;
  \item the key is approved for the asserted \texttt{clearance};
  \item the \texttt{signature} verifies over the canonical body;
  \item \texttt{clearance} dominates the host's required level (\cref{def:dominates}); and
  \item if \texttt{netAllowedHosts} is non-empty, the connected origin is a member.
\end{enumerate}
A host in a deny-by-default (high-assurance) posture \textsc{must} reject an
unattested or failing server; a permissive host \textsc{should} surface the
failure and \textsc{must not} silently treat it as success. These rules are
exactly \cref{prop:g1}, and every clause is exercised by \cref{sec:eval}.

\paragraph{Tool authorization.}
Admitting a server \textsc{must not} be read as authorizing all its tools. A host
\textsc{should} keep a per-server allow-list and \textsc{must} deny a
\texttt{tools/call} for any tool absent from it \emph{before} any network
dispatch. Per-tool \texttt{clearance} \textsc{may} be added in a future \texttt{v}.

\paragraph{Audit.}
A conforming host \textsc{should} append a tamper-evident record for every
admission decision (allow / deny / warn) and every tool-authorization denial,
carrying the resolved \texttt{signerKeyId} and \texttt{clearance} or the denial
reason (\cref{sec:atsa-errors}).

\paragraph{Trust establishment.}
How the trust root is provisioned, pinned, and sealed is deployment policy and
out of scope of the wire format. Profiles \textsc{may} anchor it in a pinned key
set (Enclawed), X.509 chains, Secure Production Identity Framework for Everyone
(SPIFFE) verifiable identity documents (SVIDs)~\cite{spiffe}, or a sigstore-style
transparency log~\cite{sigstore}; the verification rules are independent of the
choice.

\paragraph{Compatibility, versioning, negotiation.}
\texttt{v} versions the document and a verifier \textsc{must} reject versions it
does not understand. The extension is purely additive: an unextended host never
fetches the SAD and behaves exactly as today, and an unextended server never
sees the host-side allow-list; a host \textsc{must} therefore interoperate with
unattested servers under its posture, which is what makes incremental rollout
(\cref{sec:flavor}) possible.

\paragraph{Conformance.}
A conforming implementation \textsc{must} reproduce the verdicts of the test
vectors in \cref{sec:atsa-vectors}. The Enclawed \texttt{mcp-attested} extension
is a production reference implementation: its 48 hermetic tests and the LLM-driven coverage
campaign (\cref{sec:eval}) exercise every verification clause and the
tool-authorization rule against crafted \emph{and} model-generated adversarial
inputs. Wire format (\cref{sec:atsa-schema}), verification rules, error registry
(\cref{sec:atsa-errors}), well-known registration (\cref{sec:atsa-iana}), and
executable vectors (\cref{sec:atsa-vectors}) together leave no normative gap for
an implementer to fill by guesswork --- the bar an addition to the MCP
specification must clear.

\paragraph{Bootstrapping the ecosystem.} The first-party bridge fast path
(\cref{sec:flavor}) is the migration story in miniature: today's public MCP
servers do not publish attestation documents, so the host's trust in them is
mediated by a \emph{locally} signed bridge manifest instead. As servers begin
publishing well-known attestation documents, the same host gate verifies them
directly and the bridge shim falls away --- exactly how a well-known-URI
extension is meant to roll out incrementally.

\section{Composition with runtime drift monitoring}
\label{sec:drift-composition}

ATSA establishes trust strictly at the point of admission: it verifies an
unforgeable server identity and evaluates a pinned signer key before any
network dispatch. This is a point-in-time guarantee that answers whether a
host is authorized to communicate with a specific server at a designated
sensitivity level at the moment of connection. By design, the wire format
does not address whether a tool's internal behavior or exposed schema drifts
over the lifetime of an active session. An admitted server may pass
verification successfully but subsequently mutate its tool surface: a tool
declared as read-only during an initial handshake could later advertise
mutating side effects, inject new sensitive-data parameters (e.g., personally
identifiable information), escalate network externalities, or trigger dynamic
schema modifications between polling intervals. Because the underlying server
identity remains unchanged, the cryptographic admission guarantee holds, but
the capability surface has drifted. This operational boundary is where a
downstream runtime drift-monitoring layer composes with ATSA.

\paragraph{What the drift layer needs from ATSA.}
A continuous-monitoring framework requires a stable, cryptographically
verified anchor to bind its behavioral baselines to. ATSA provides this
structural dependency via two primitives:
\begin{enumerate}[leftmargin=1.4em,itemsep=2pt]
  \item \textbf{Verified server identity (\texttt{id}).} The unforgeable
        identity string that the runtime monitor uses as the primary
        indexing key for its tool-schema baseline. Without this
        cryptographic constraint, a drift monitor can be spoofed into
        baselining an adversarial substitute server.
  \item \textbf{Pinned signer key (\texttt{signerKeyId}).} The cryptographic
        guarantee that the entity producing tool definitions at execution
        time matches the entity that was admitted. This ensures an attacker
        cannot bypass drift telemetry by impersonating an admitted origin.
\end{enumerate}
The downstream monitor anchors each tool baseline to the unique tuple
$(\texttt{id},\texttt{toolName})$ at the first observation post-admission,
checking subsequent \texttt{tools/list} or \texttt{tools/call} envelopes
against that baseline.

\paragraph{Explicit non-requirements of the wire format.}
To prevent protocol scope creep and maintain structural minimalism, three
boundaries are enforced. \emph{State isolation:} the wire format
\textsc{must not} carry drift state, baseline histories, or schema-versioning
metadata; baseline persistence and delta evaluation are strictly
implementation concerns of the runtime monitor. \emph{Telemetry isolation:}
the wire format is not responsible for detecting or signaling structural
mutations --- it provides the authenticated identity context; detecting
change is the monitor's responsibility. \emph{Frequency decoupling:} ATSA
is \textsc{not required} to execute full cryptographic re-attestation on
every individual message frame; admission remains an edge gate, and
continuous runtime verification lives in the downstream monitoring layer.

\paragraph{Systemic synergy.}
Admission control without drift monitoring risks trusting a modified
capability surface indefinitely. Conversely, drift monitoring without a
verified admission anchor operates in the dark --- its baselines remain
completely vulnerable to origin spoofing. Composed, the two layers provide
a complete security posture: ATSA guarantees \emph{who}, while the drift
layer guarantees that the \emph{who} hasn't silently changed \emph{what}.

\section{Related work}

\paragraph{Guardrails for LLM applications.} Toolkits such as NeMo
Guardrails~\cite{nemo-guardrails} police the \emph{content} of model inputs and
outputs with programmable rails. They are complementary and downstream of this
work: a content rail does not decide \emph{which server} a tool call may reach
or \emph{at what clearance}; it inspects the message once the connection is
assumed. Enclawed pairs admission control (this paper) with a multi-modal egress
reference monitor~\cite{metere2026covert} for the content side.

\paragraph{Workload and service identity.} SPIFFE/SVID~\cite{spiffe} and mutual
TLS give services cryptographic \emph{identities}; our clearance assertion adds
an orthogonal \emph{authorization-to-a-sensitivity-level} attribute and a
host-side admission ordering on top of whatever identity layer is present. A
SPIFFE deployment is a natural trust-root provider for the standardized profile.

\paragraph{Software-supply-chain attestation.} The Update Framework (TUF)~\cite{tuf},
in-toto~\cite{in-toto}, and sigstore~\cite{sigstore} sign and verify
\emph{artifacts} and build provenance; the clearance assertion borrows their
offline-signing and pinned-root discipline but applies it to a \emph{live
endpoint's authorization}, verified at connection time rather than install
time.

\paragraph{Information-flow and access control.} The clearance lattice is a
direct application of Bell--LaPadula~\cite{bell-lapadula} and the
least-privilege principle~\cite{saltzer-schroeder} to the MCP admission point;
the confused-deputy framing~\cite{confused-deputy} is exactly the risk a
self-asserted, wholesale-trusted tool list creates. The companion skills
paper~\cite{metere2026skills} provides the capability-containment layer that
governs behavior \emph{after} admission.

\section{Limitations and future work}
\label{sec:limits}

Several directions remain open. \textbf{Per-tool clearance.} The assertion binds
a level to a server, not to individual tools; letting one server expose a
\texttt{PUBLIC} search tool beside a \texttt{RESTRICTED} send tool is a natural
refinement that the registry's per-server allow-list already half-anticipates.
\textbf{Per-assertion freshness and revocation.} Beyond the per-signer
\texttt{notAfter} the verifier enforces, a short-lived \texttt{notBefore}/expiry
on the assertion itself and an online, Online-Certificate-Status-Protocol (OCSP)-style revocation channel for
compromised signer keys would tighten the freshness story. \textbf{Standardization.}
The normative extension of \cref{sec:atsa} is specified but not yet submitted to
the MCP maintainers; reconciling its \texttt{clearance} field with MCP's evolving
OAuth profile and capability-negotiation handshake is the concrete next step.

\section{Conclusions}

MCP left trust to the deployment, and that gap is exploitable everywhere --- a
prompt-injected model can drive a destructive tool on any server it reaches. A
lone operator may absorb that risk; a regulated one cannot, and without an
admission record cannot even audit it. We showed the missing trust layer can be
added \emph{entirely above} MCP, without changing a single message, and backed
each guarantee (authenticated admission, tool least-privilege, auditability,
tamper-resistance) with crafted and LLM-generated adversarial tests.

\paragraph{Why this belongs in the MCP specification, not in $N$ vendor forks.}
Several forces argue for standardizing admission at the spec level rather than
leaving it to each deployment:
\begin{itemize}[leftmargin=1.4em,itemsep=2pt]
  \item \textbf{Trust does not compose across dialects.} If every host and
        server invents its own attestation scheme, none interoperate and the
        ecosystem's security ceiling is set by its least-attested member. One
        standardized shape --- a single well-known document and a single
        verification ordering --- is what makes ``this server is attested'' a
        portable claim instead of a private convention.
  \item \textbf{The gap is intrinsic to MCP, so the fix belongs at MCP's layer.}
        MCP deliberately omitted trust; pushing that omission onto deployments
        does not remove it, it fragments it --- forcing every regulated adopter
        to reinvent admission control, usually worse and never interoperably.
  \item \textbf{Network effects need a Schelling point.} A well-known URI and a
        capability handshake only pay off once both sides converge on one
        format. Public MCP servers publish no attestation today precisely
        because there is no standard to publish \emph{against}; a spec creates
        the target, after which adoption is incremental and self-reinforcing
        (the bridge path of \cref{sec:flavor} is that rollout in miniature).
  \item \textbf{Standardizing costs non-adopters nothing.} Because the extension
        is purely additive it can be \textsc{optional} in the spec yet uniform
        in shape: an unextended host or server is wholly unaffected, while
        adopters gain interoperable admission. There is no compatibility tax to
        set against the benefit.
  \item \textbf{It is cheaper now than later.} MCP is young; agentic AI on
        regulated data is not waiting. Retrofitting a trust layer after a large
        installed base has hard-coded ``connect and trust'' is far costlier than
        reserving the well-known name and the document shape today.
\end{itemize}

\noindent We have also removed the usual reasons to defer. The extension is given
normatively (\cref{sec:atsa}) with a JSON Schema, an error registry, a
well-known-URI registration, and machine-checkable conformance vectors, and it
ships with a reference implementation whose adversarial evaluation drives every
verification clause with tens of thousands of model-generated inputs.
``Incomplete,'' ``unproven,'' and ``it forks the protocol'' are no longer
available objections. What remains is a decision the specification's maintainers
are uniquely positioned to make: to let agentic MCP reach regulated environments
by adopting attested tool-server admission as a standard, optional, additive
part of the protocol --- rather than leaving each adopter to rebuild it alone.

\appendix

\section{SAD JSON Schema}
\label{sec:atsa-schema}

The Server Attestation Document conforms to the following JSON Schema
(Draft 2020-12). \texttt{signature} covers the canonical body
(\cref{def:canonical}) of all other fields.

\begin{lstlisting}[language=,caption={Server Attestation Document schema.},captionpos=b]
{
  "$schema": "https://json-schema.org/draft/2020-12/schema",
  "title": "MCP Server Attestation Document",
  "type": "object",
  "required": ["v","id","publisher","version","clearance",
               "capabilities","signerKeyId","signature"],
  "additionalProperties": false,
  "properties": {
    "v":            { "const": 1 },
    "id":           { "type": "string", "minLength": 1 },
    "publisher":    { "type": "string", "minLength": 1 },
    "version":      { "type": "string", "minLength": 1 },
    "clearance":    { "type": "string", "minLength": 1,
                      "description": "canonical name or alias in the active scheme" },
    "capabilities": { "type": "array", "items": { "type": "string" },
                      "contains": { "const": "mcp-server" } },
    "signerKeyId":  { "type": "string", "minLength": 1 },
    "netAllowedHosts": { "type": "array", "items": { "type": "string" },
                      "description": "optional; non-empty binds the SAD to these origins" },
    "verification": { "type": "string" },
    "signature":    { "type": "string", "contentEncoding": "base64" }
  }
}
\end{lstlisting}

\section{Error registry}
\label{sec:atsa-errors}

When a host surfaces an attestation rejection in-band it \textsc{should} carry a
machine-readable reason in \texttt{data.reason}. The reference reasons, one per
verification clause of \cref{prop:g1}:

\begin{center}\small
\begin{tabular}{lll}
\toprule
\texttt{data.reason} & clause & meaning \\
\midrule
\texttt{not\_mcp\_server}   & (a) & \texttt{capabilities} lacks \texttt{"mcp-server"} \\
\texttt{unsigned}           & (b) & \texttt{signerKeyId}/\texttt{signature} absent \\
\texttt{signer\_not\_trusted} & (c) & signer not in trust root \\
\texttt{signer\_expired}    & (d) & signer \texttt{notAfter} elapsed \\
\texttt{signer\_not\_approved} & (e) & signer not approved for level \\
\texttt{bad\_signature}     & (f) & signature does not verify \\
\texttt{below\_required}    & (g) & level below the host's requirement \\
\texttt{host\_not\_bound}   & (h) & origin not in \texttt{netAllowedHosts} \\
\texttt{tool\_not\_admitted} & --- & \texttt{tools/call} for a non-allowlisted tool \\
\bottomrule
\end{tabular}
\end{center}

\section{Well-known URI registration}
\label{sec:atsa-iana}

Per RFC~8615~\cite{rfc8615}, the registration template:
\begin{center}\small
\begin{tabular}{ll}
\toprule
URI suffix & \texttt{mcp-attestation} \\
Change controller & MCP maintainers \\
Specification document & this extension (\cref{sec:atsa}) \\
Status & permanent (proposed) \\
Media type & \texttt{application/json} \\
\bottomrule
\end{tabular}
\end{center}
\noindent The Enclawed profile additionally serves the same document at
\wellknown\ for backward compatibility with deployed clients.

\section{Conformance test vectors}
\label{sec:atsa-vectors}

A conforming verifier \textsc{must} reproduce the verdicts below for a trust
root containing one signer \texttt{S} (Ed25519) approved up to
\texttt{restricted-plus}, required level \texttt{restricted-plus}. Each vector is
a single-mutation derivation of vector~1, mirroring the reference suite
(\cref{sec:eval}); ``ADMIT/DENY($x$)'' names the deciding clause of
\cref{prop:g1}.

\begin{center}\small
\begin{tabular}{clp{0.40\linewidth}}
\toprule
\# & verdict & vector (relative to a valid baseline signed by \texttt{S}) \\
\midrule
1 & \textbf{ADMIT}     & valid baseline: \texttt{clearance=restricted-plus}, signed by \texttt{S}, \texttt{netAllowedHosts=[]} \\
2 & DENY (a)           & \texttt{capabilities} = \texttt{["tool.invoke"]} \\
3 & DENY (b)           & \texttt{signature} removed \\
4 & DENY (c)           & \texttt{signerKeyId} = \texttt{"unknown"} \\
5 & DENY (d)           & \texttt{S.notAfter} in the past \\
6 & DENY (e)           & \texttt{S} approved only to \texttt{internal} \\
7 & DENY (f)           & one byte of \texttt{signature} flipped \\
8 & DENY (f)           & \texttt{clearance} raised to \texttt{restricted-plus} after signing at \texttt{internal} \\
9 & DENY (g)           & \texttt{clearance=internal}, required \texttt{restricted-plus} \\
10 & DENY (h)          & \texttt{netAllowedHosts=["a.example"]}, served from \texttt{b.example} \\
11 & ADMIT             & \texttt{netAllowedHosts=["a.example"]}, served from \texttt{a.example} \\
\bottomrule
\end{tabular}
\end{center}
\noindent These eleven vectors are the \texttt{T2} cases of the reference
implementation's adversarial suite and pass against it; the LLM-driven campaign
(\cref{sec:eval}) extends vectors 2--10 to tens of thousands of model-generated
instances (27{,}025 tool names, 14{,}378 forged assertions).

\section*{Acknowledgements}

The author thanks \textbf{Maaz (@maaz-interlock)} of Interlock for critical
contributions to the architectural framing of the admission/runtime boundary
and for drafting the normative composition language for downstream continuous
drift-monitoring frameworks (\cref{sec:drift-composition}); and
\textbf{Christopher Hopley and the AlgoVoi team} for deploying the
\emph{Agent-Trust-Bench} production adversarial suite, which empirically
validated the runtime threat taxonomy ATSA closes and supplied the
defender-side baseline data referenced in \cref{sec:threat,sec:eval}. Their
contributions are reflected in the corresponding sections of the companion
SEP (\texttt{modelcontextprotocol/modelcontextprotocol\#2809}).

\bibliographystyle{plainnat}
\bibliography{refs}

\end{document}